\documentclass{article}  
\usepackage{spadre2008}
\usepackage[dvips]{graphicx}
\usepackage[hang,small]{caption}
\frompage{000} \topage{000}                                              
\def\rightmark{Recent PHENIX Spin Results}
\def\leftmark{A. Morreale for the PHENIX Collaboration}
 
\title{Recent Spin Results from the PHENIX Detector at RHIC} 
\authors{
{Astrid Morreale$^1$ for the PHENIX Collaboration$^2$ %
}\\[2.812mm]
{\normalsize
\hspace*{-8pt}$^1$ University of California Riverside,$^2$ http://www.phenix.bnl.gov \\[0.2ex] 
}}
\abstract{The PHENIX Experiment on the Relativistic Heavy Ion Collider (RHIC) with its use of beams of polarized protons, provides a unique environment of hard scattering between
gluons and quarks complementary to that provided by deep inelastic scattering (DIS).
Polarized proton-proton collisions can directly probe the polarized gluon and
anti-quark distributions as the collisions couple the color charges of the participants. We will give a brief
overview of the PHENIX Spin Program and we will report recent results of the many probes accessible to the
PHENIX experiment.}

\keyword{proton, spin, gluon, orbital angular momentum, SSA, delta g} 
\PACS{14.20Dh, 13.85.Ni,25.40.Ep, 13.88.+e, 14.70.Dj}
 
\begin{document}

\renewcommand{\topfraction}{0.95}
\renewcommand{\textfraction}{0.05}
\maketitle
\setcounter{page}{1}
\section{Polarized RHIC and PHENIX}
The spin of a particle is a quantum mechanical attribute that probes underlying theoretical structures deeply: in quantum chromodynamics (QCD) the spin of the proton plays a central role in the understanding of the nucleon structure. A surprising result was found in the late 1980's by the European Muon Collaboration (EMC) at CERN that the spin of the quarks contributed a very small fraction to the proton's spin.
The original EMC publication[1] that triggered the {\it spin crisis} has since resulted in a large theoretical and experimental effort to find the pieces to the proton's spin puzzle, mainly, what role gluons, sea quarks and orbital angular
momentum (OAM) play. This can be summarized in the helicity sum rule (Eq.\ref{eq1}) where $\Delta$(u, d, s, G) are the probabilities of finding a q, $\bar{q}$ or gluon with spin parallel or anti-parallel to the spin of the nucleon and L is the OAM of the parton.
 
\begin{equation}
 S_{z}= \frac{1}{2} =\frac{1}{2} (\Delta u +\Delta d +\Delta s) + L_{q} +\Delta G +L_{G} \, 
 \label{eq1}
\end{equation} 

The RHIC spin program complements the work done in DIS by making use of strongly-interacting polarized quark and gluonic probes which are sensitive to the gluon polarization $\Delta g$, a major emphasis of RHIC-spin[2]. The transverse spin structure of
the proton is also being explored with measurements sensitive to the sivers effect, transversity and the collins effect. Future running at $\sqrt{s} = 500 GeV$ will focus at disentagling quark spin-flavor separation in W-Boson production thus measuring the flavor asymmetry of the polarized antiquark sea [2].

Polarizing protons is not a trivial task; maintaining proton polarization is a challenge due to the proton's large anomalous magnetic moment. The novel use of RHIC's siberian snakes, avoids major depolarizing spin resonances in the beams and a stable spin direction can be obtained.(For details see \cite{bib3})

The PHENIX detector located at the 8 o'clock position at RHIC has two central arms with pseudorapidity acceptance of $|\eta | < 0.35$. These are equipped with fine-grained calorimetry 100 times finer than previous collider detectors, making particle identification excellent, the granularity of the electromagnetic calorimeter (EMCal) is $\Delta{\eta} \times \Delta{\phi}$ = $0.01 \times 0.01.$ \cite{bib4}. Triggering in the central arms allow us to select high $p_{T}$ $\gamma$, $e^{\pm}$ and $\pi^{\pm}$. The PHENIX muon arms cover $1.2 < \eta <2.4 $,  they surround the beams and include $\mu^{\pm}$ identifiers, tracking stations and iron sheets with detectors in the gaps in each sheet. 
 
\section{The Hunt for $\Delta g$}
Measurement of $\Delta g$ is one of the main goals of PHENIX-spin . Under factorization, a differential cross section can be written as the convolution of a parton density function (pdf) and a hard scattering process; this along with universality of pdf's and fragmentation functions (FF, $D_{h}$) allows for separation of long and short distance contributions in the cross sections.  In practice, what are measured are asymmetries: the ratio of the polarized to unpolarized cross sections (Eq. 2). Asymmetries give an elegant way of accessing parton information by counting observed particle yields in different helicity states of incident protons ( $++, --$ vs $+-, -+.$) normalized by the polarization in each beam ($P_{B,Y}$) 
 
\begin{eqnarray*}
A_{LL} & = &  \frac{\displaystyle\sum_{a,b,c={\rm q}, \bar{\rm q}, {\rm g}} 
\Delta f_{a}\otimes \Delta f_{b} \otimes \Delta \hat{\sigma}\otimes D_{h/c}} 
{\displaystyle\sum_{a,b,c={\rm q}, \bar{\rm q}, {\rm g}} f_{a} 
\otimes f_{b} \otimes \hat{\sigma} 
\otimes D_{h/c}} = \frac{\sigma_{++} -\sigma_{+-}}{\sigma_{++}+\sigma_{+-}} \,, 
\nonumber\\
 A_{LL} & = & \frac{1}{P_{B}P_{Y}} \frac{N^{++}-RN^{+-}}{N^{++}+RN^{+-}}\,, 
\qquad\qquad  
R = \frac{L_{++}}{L_{+-}} \,.
\label{eq2}
\end{eqnarray*}

\subsection{Latest $A_{LL}$ results}

\begin{figure}[h]
\centering  
\includegraphics[scale=0.185]{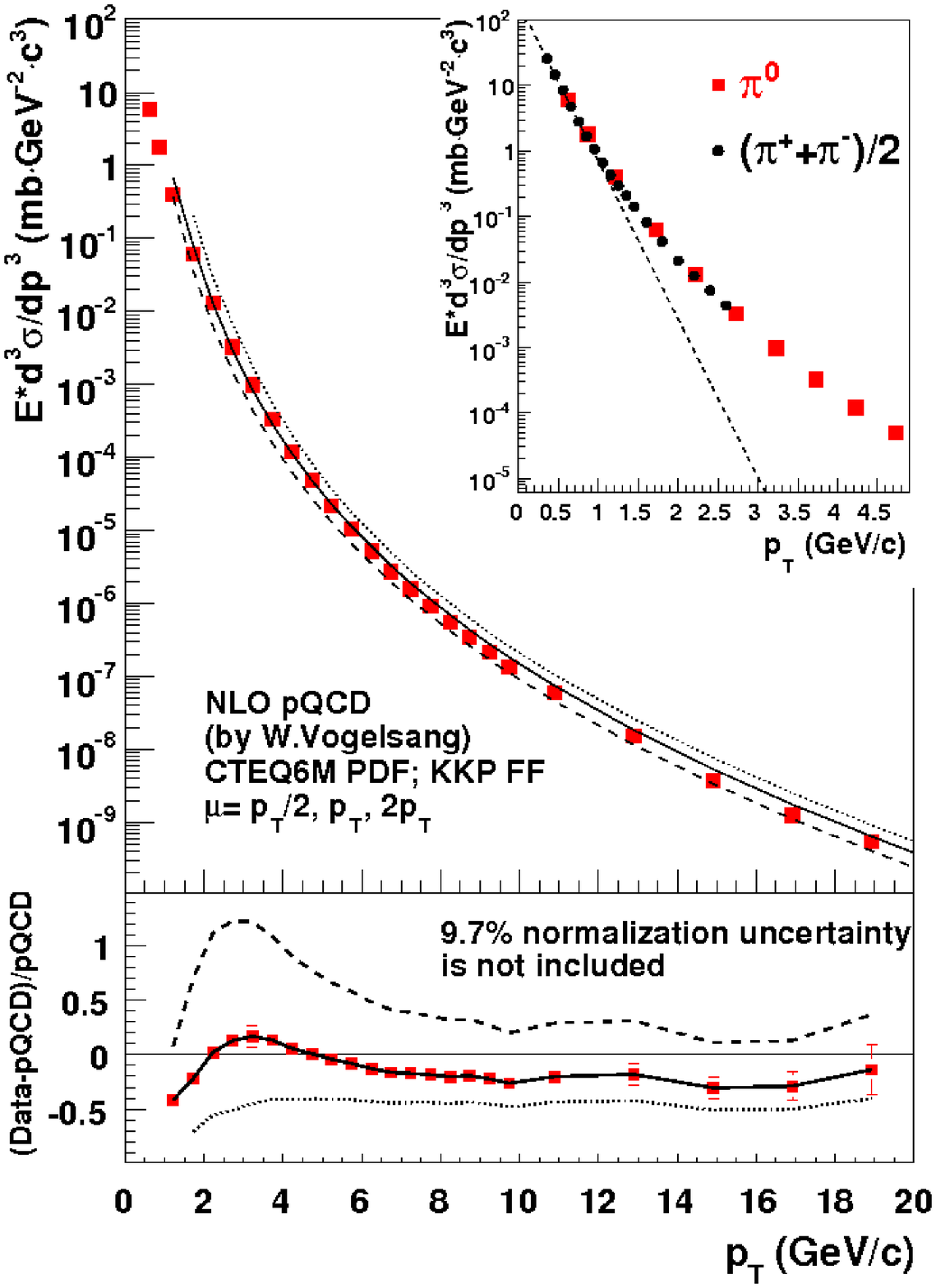}\includegraphics[scale=0.175]{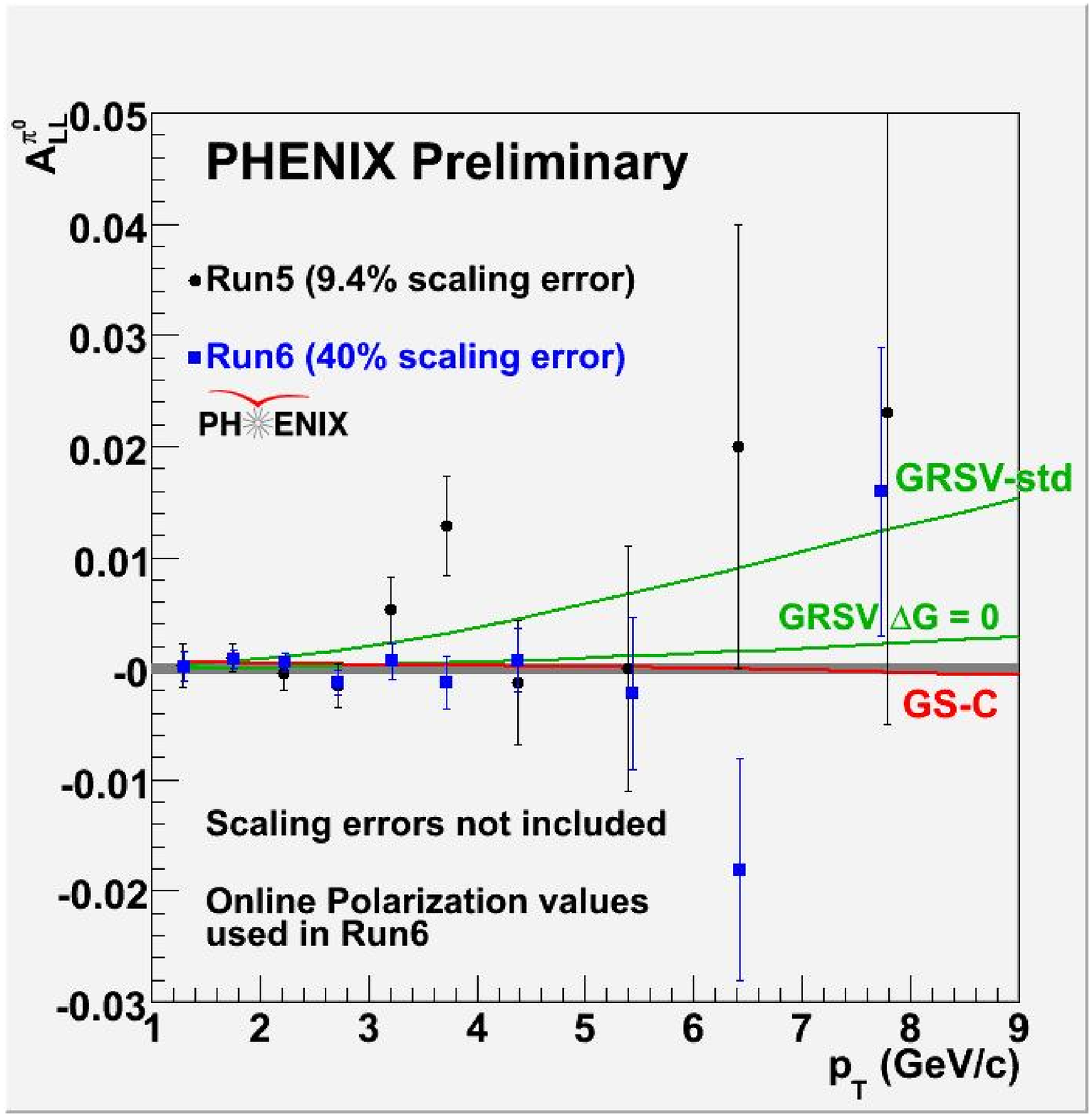}
\caption{(left)$\pi^{0}(p_{T})$ cross section. (right)$\pi^{0} A_{LL}(P_{T})$ Curves based on  GRSV(6), GS-C(7)}
\label{fig1}
\end{figure}

Measuring $A_{LL}$ in certain final states is a valuable tool to measure polarized gluon distribution functions in the proton. The most accurately way to do so is to study those processes which can be calculated in the framework of pQCD. PHENIX unpolarized $\pi^{0}$ \cite{bib11}(Fig. \ref{fig1}) and prompt $\gamma$ production cross sections at mid rapidity  have shown that the next to leading order (NLO) perturbative calculation describe the data well at RHIC energies. 

\begin{figure}[h]
\centering
\includegraphics[scale=0.165]{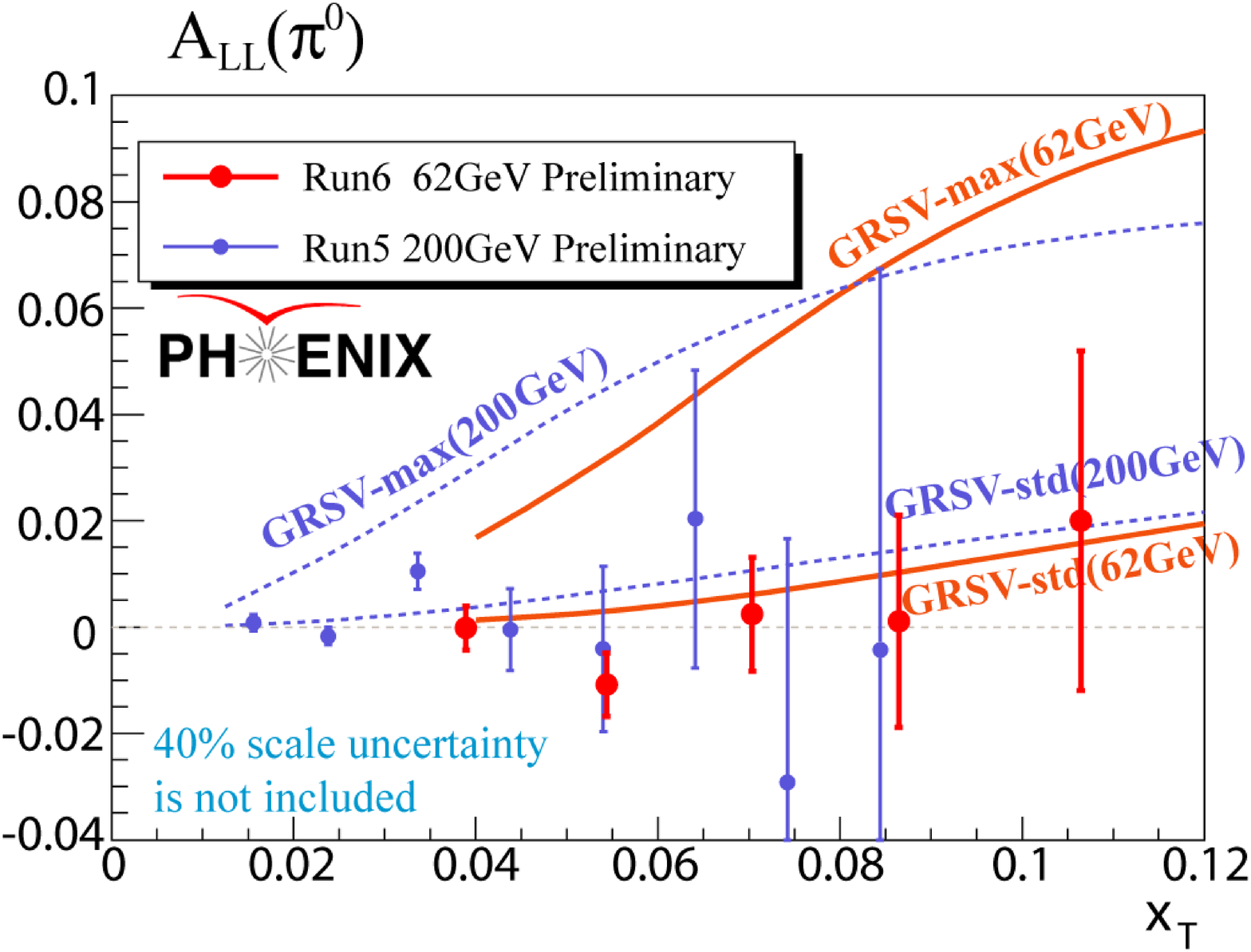}\includegraphics[scale=0.275]{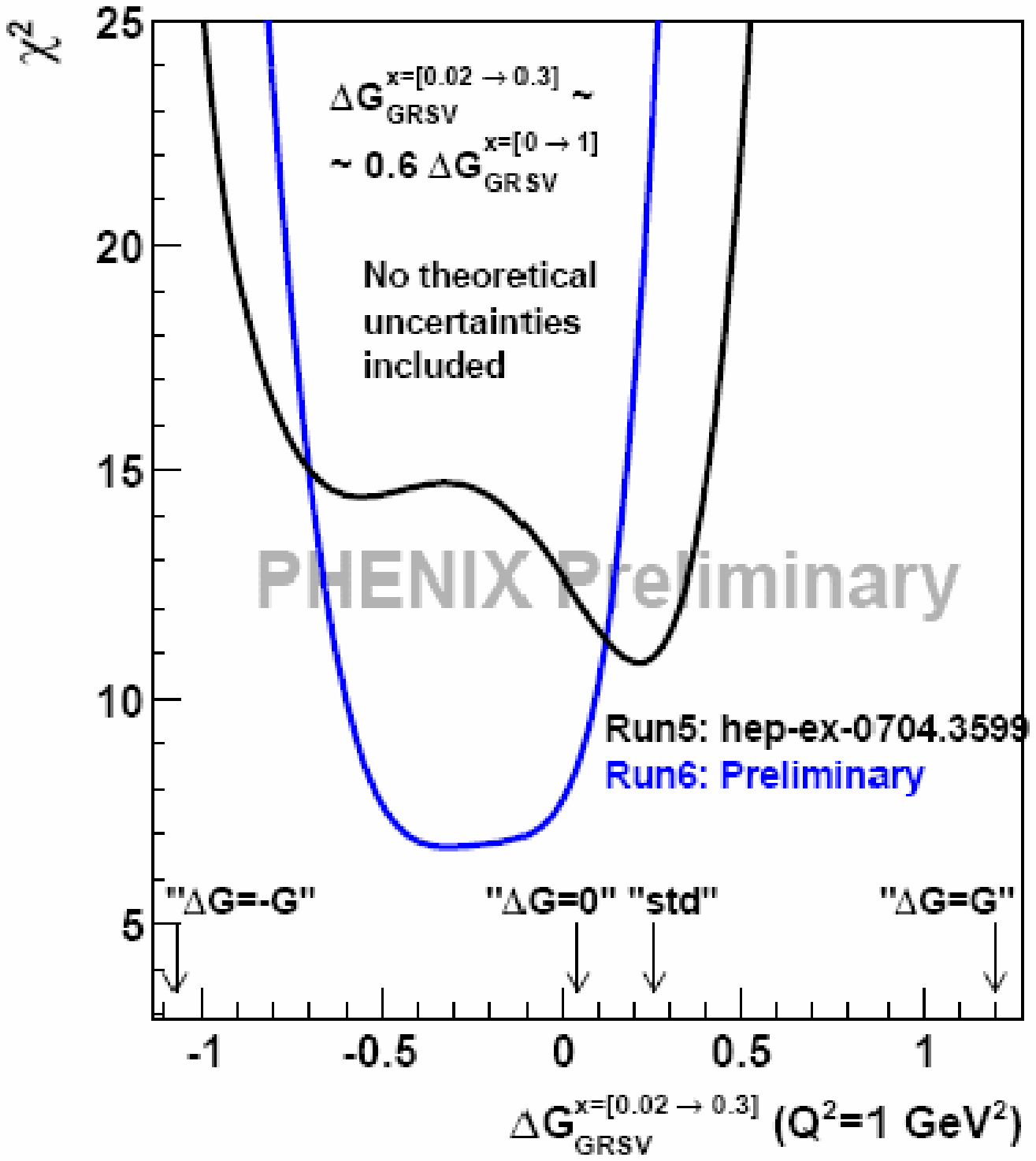}

\caption{(left)$\pi^{0} A_{LL}(x_{T})$ at $\sqrt{s}$ =62.4 GeV and 200GeV.
         (right) $\chi^{2}$ derived by comparing the measurement to a range of 
         solutions provided by GRSV, for which the value of the integral  
          $\Delta G$ was constrained and the lepton scattering data refit.}

\label{fig2}
\end{figure}

A variety of probes have been measured at PHENIX , $\pi^{0}$'s \cite{bib11}(Fig. \ref{fig1})being the most mature of the probes with measurements at $\sqrt{s}$ = 200 GeV, and 62.4 GeV (Fig. \ref{fig2}). The latest measured asymmetries have now been included for the first time in a NLO global analysis. \cite{bib5} $\pi^{0}$'s  have significantly constrained $\Delta$g, however, large uncertainties remain and sign information of $\Delta$g is still unknown. 

\begin{figure}[h]
\centering
\includegraphics[scale=0.231]{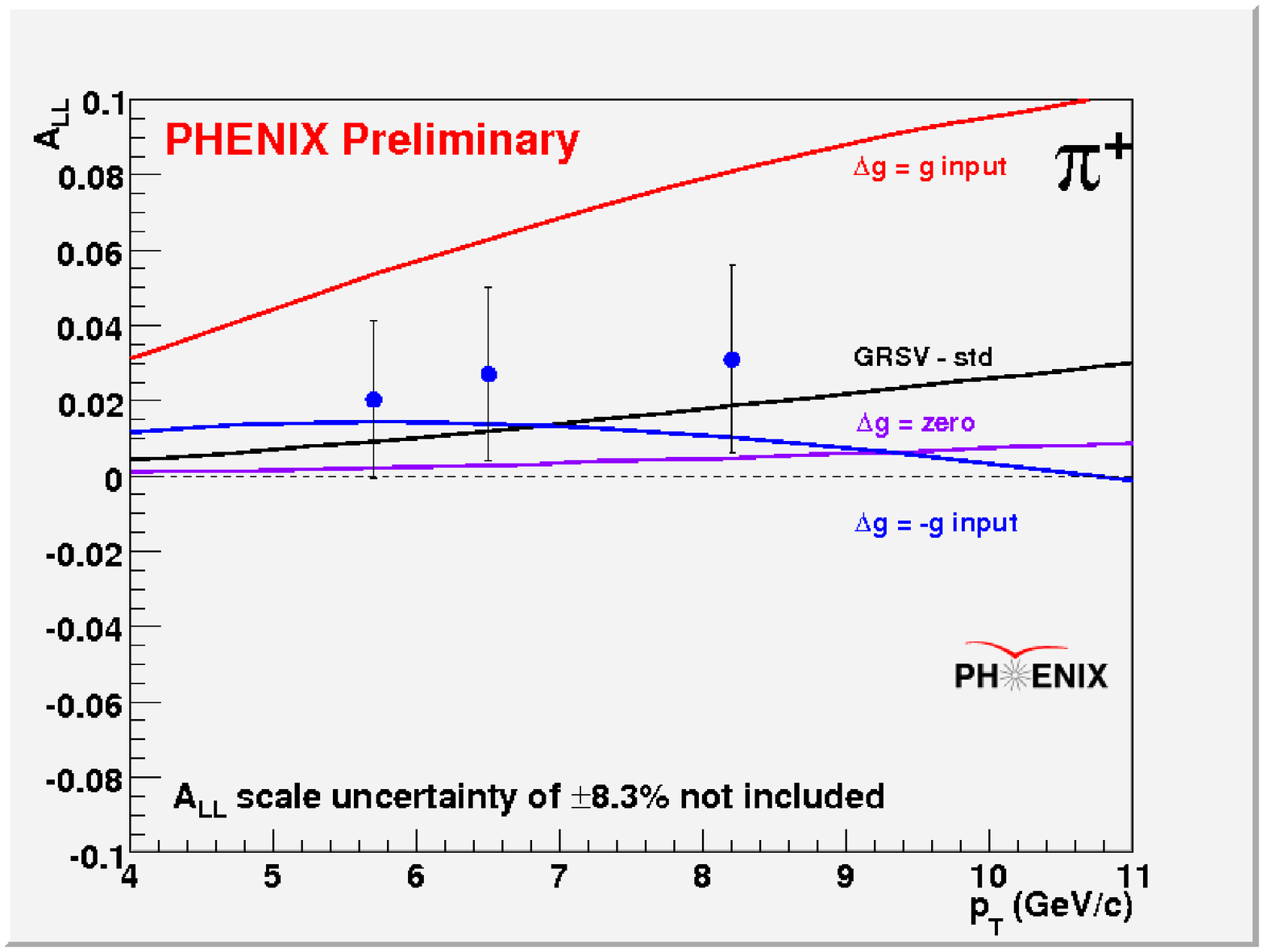}\includegraphics[scale=0.241]{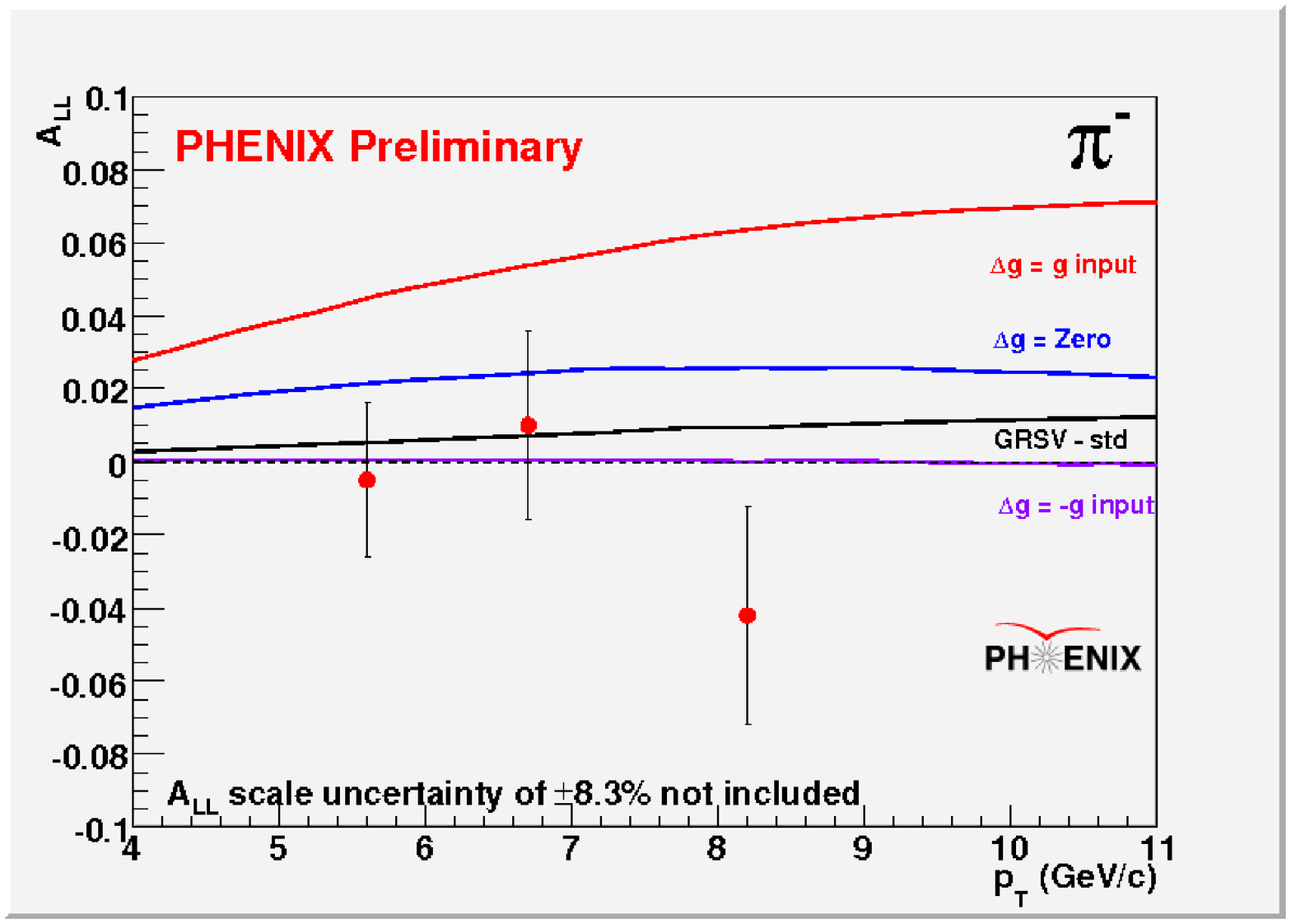}
\caption{(left) $\pi^{+} A_{LL}(P_{T})$, (right) $\pi^{-} A_{LL}(P_{T})$. $\sqrt{s}$=200GeV. }
\label{fig3}
\end{figure}

\renewcommand{\topfraction}{0.95}
\renewcommand{\textfraction}{0.05}
 Charged pions are a model independent probe which are sensitive to the sign and magnitude of $\Delta$g:
quark-gluon (qg) scattering dominates mid-rapidity pion production at RHIC at
transverse momenta above 5 GeV/c. Preferential fragmentation of up quarks (u)
to $\pi^{+}$, and down quarks(d) to $\pi^{-}$, leads to the dominance of
u-g, and d-g  contributions. This dominance of u or d combined with the
different signs of their polarized distributions translates into
asymmetry differences for the different species $\pi^{+}$, $\pi^{0}$, and $\pi^{-}$ (Fig.\ref{fig5}) that depend on the sign of $\Delta$g.
For example, a positive $\Delta$g could be indicated by an order of $\pi$ asymmetries, i.e: 
$A_{LL}(\pi^{+})> A_{LL}(\pi^{0}) > A_{LL}(\pi^{-})$, and viceversa for a negative contribution.
The latest measured $\pi^{\pm} A_{LL}$ (2006)(Fig.\ref{fig3}) have been compared with a new set of theoretical calculations. These for the first time use charge separated $\pi^{\pm}$ data for $D_{\pi^{\pm}}$ extraction. $\pi^{\pm}$ are expected to be players in future global analyses of $\Delta$g.

\begin{figure}[htb]
\centering 
\includegraphics[scale=0.075]{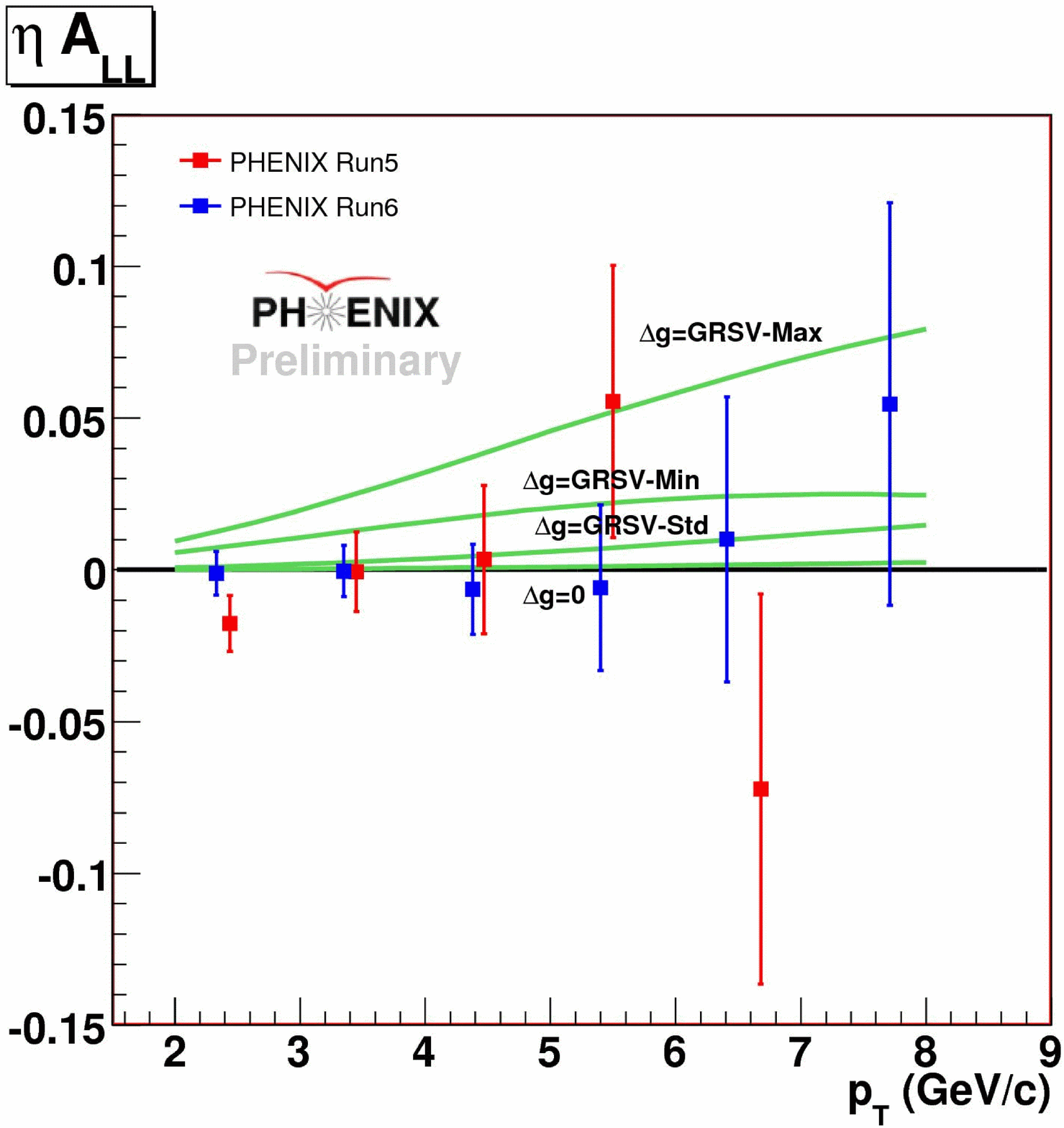}\includegraphics[scale=0.220]{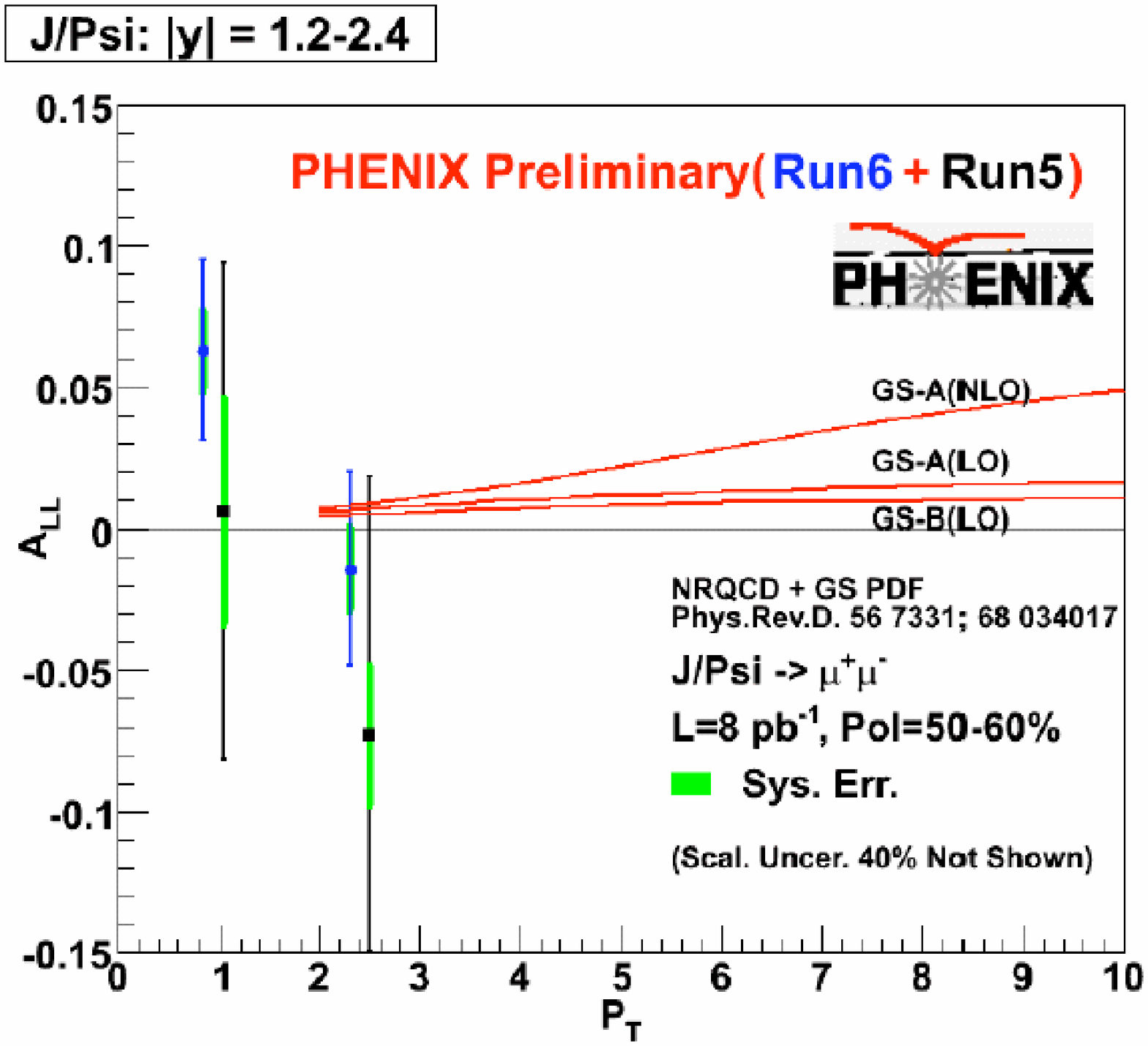}
\caption{(left)$\eta$ $A_{LL}(P_{T})$ at $sqrt{s}$200GeV. (right)$A_{LL}(P_{T})$ of $J/\psi \longrightarrow \mu^{+}\mu^{-}$, }
\label{fig4}
\end{figure} 
Particle cluster asymmetries (Fig. \ref{fig5}), as well as $A_{LL}$ of $\eta$  have been calculated. The recent preliminary extraction of $\eta$'s $D_{\eta}$, has allowed for $A_{LL}$ theory comparisons. $D_{\eta}$ shows a slight enhanced sensitivity to qg when compared to $\pi^{0}$ . Observation of difference in asymmetries could help disentangle the contributions from the different quarks and gluons.(Fig.\ref{fig4})
Rare channels measured at PHENIX include: $A_{LL}$ of $\mu^{\pm}$ coming from $J/\psi$ (Fig.\ref{fig4}) production, $e^{\pm} A_{LL}$ coming from heavy quarks,(Fig. \ref{fig6}) and prompt $\gamma$ (Fig. \ref{fig6}.) Prompt $\gamma$ are a clean elegant probe dominated by qg compton and thus can give a better access to $\Delta$g, however as is the case for rare probes, these measurements require high luminosities, not yet achieved at RHIC.
\renewcommand{\topfraction}{0.95}
\renewcommand{\textfraction}{0.05}
\begin{figure}[h]
\centering
\includegraphics[scale=0.231]{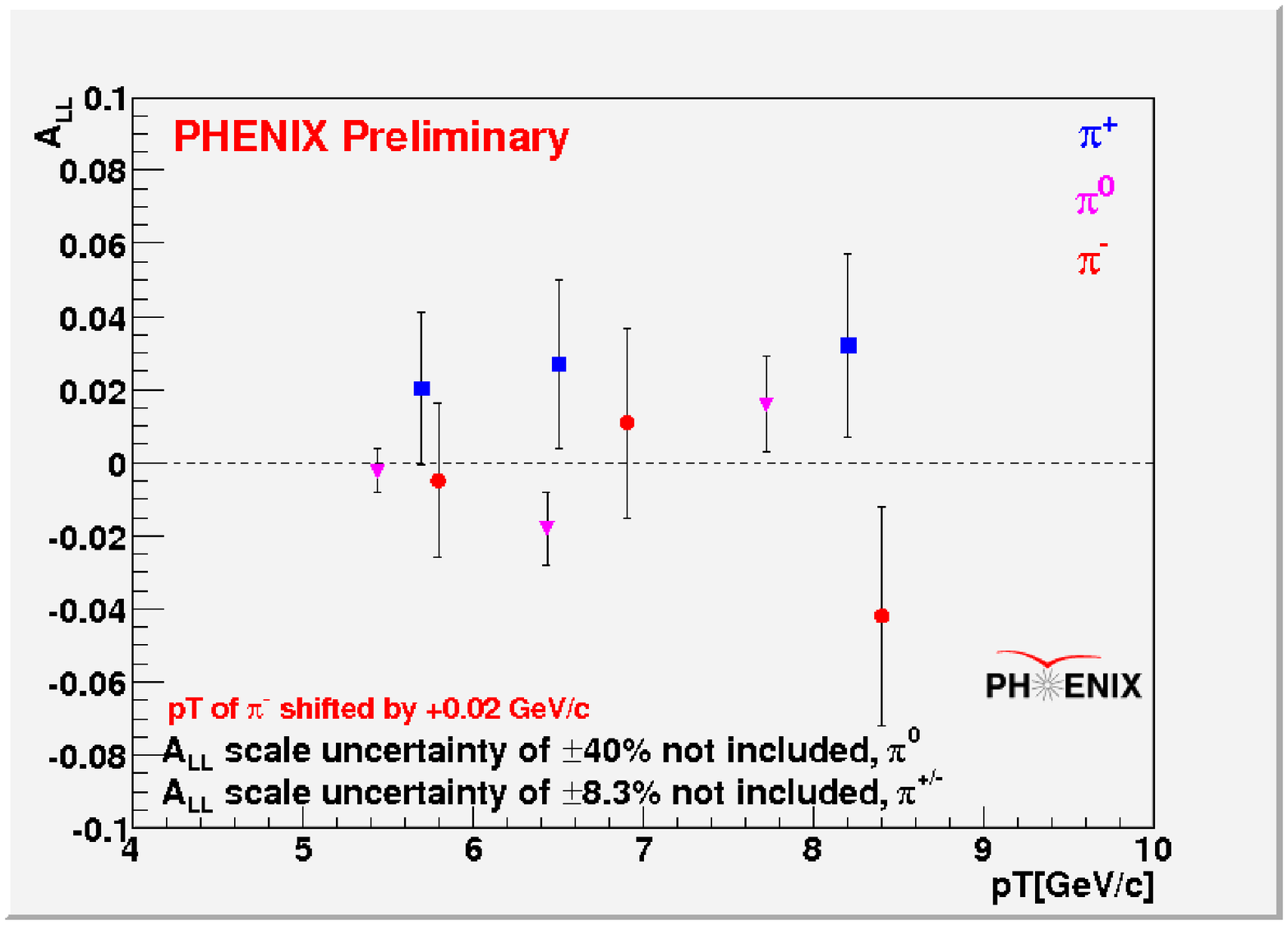}\includegraphics[scale=0.115]{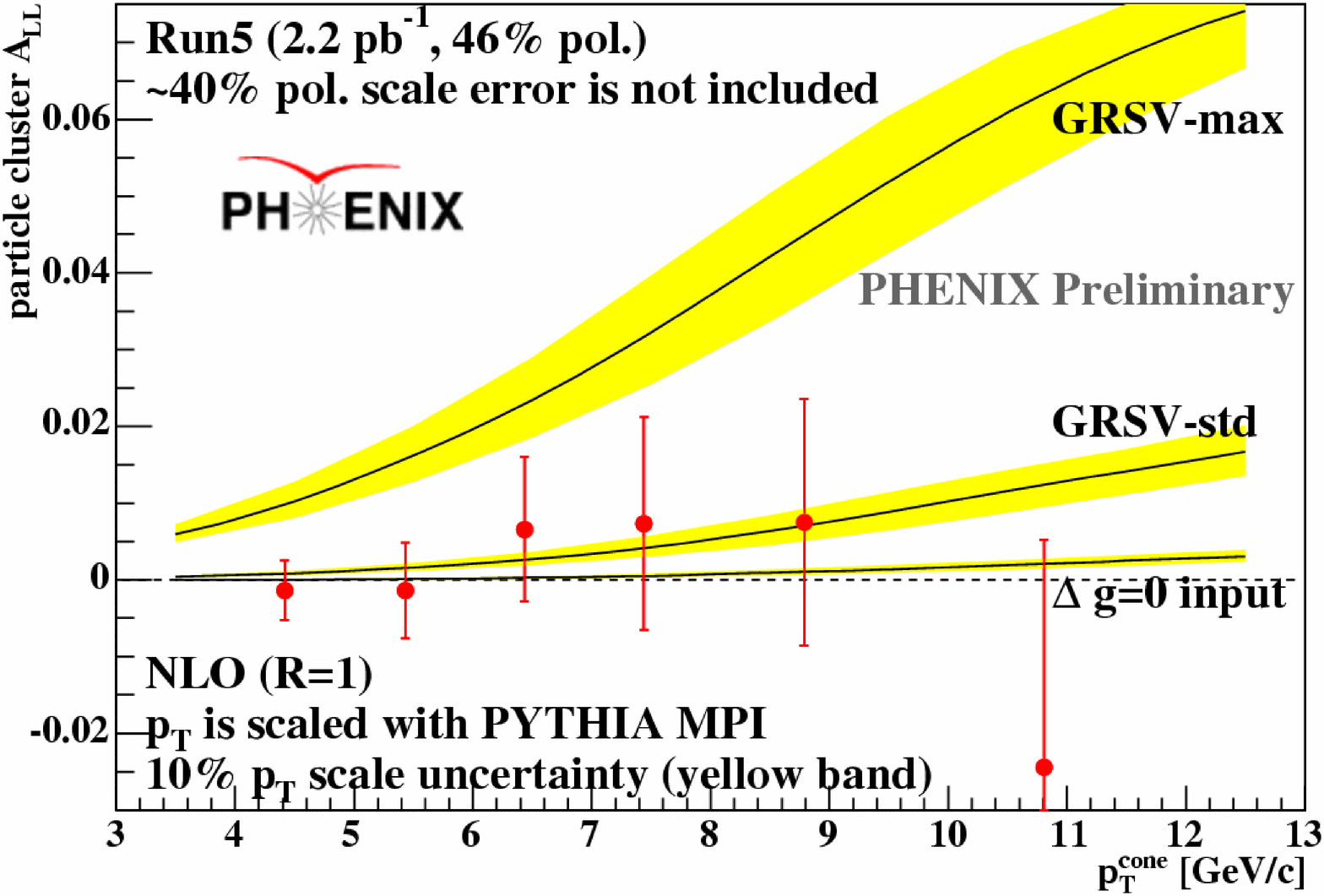}
\caption{(left) $\pi^{\pm , 0}$'s $A_{LL}(p_{T})$ at $\sqrt{s}=200GeV$. (right)Particle cluster $A_{LL}$ correlated with $\pi^{0}$}
\label{fig5}
\end{figure} 

\begin{figure}[h]
\centering 
\includegraphics[scale=0.182]{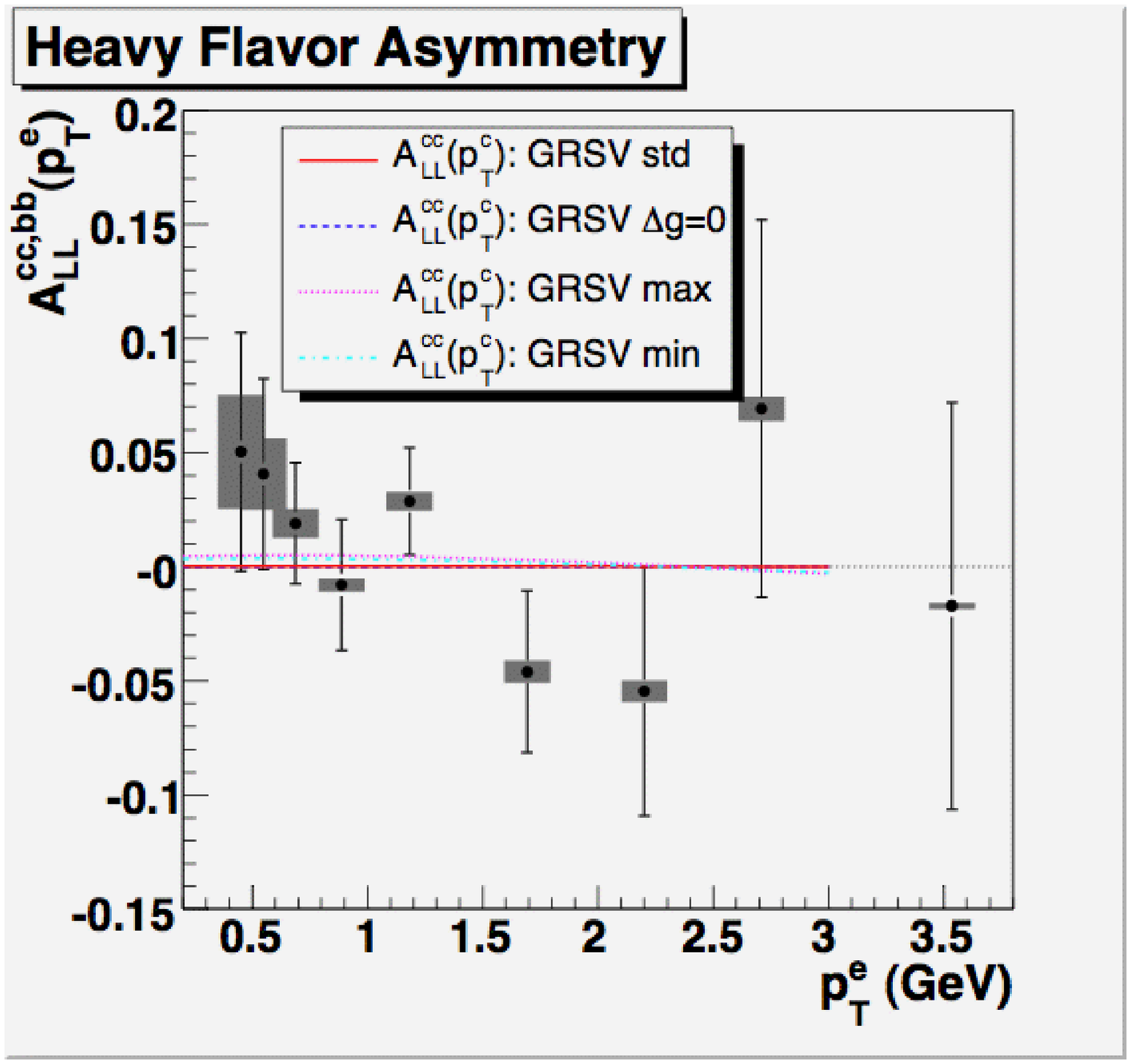}\includegraphics[scale=0.16]{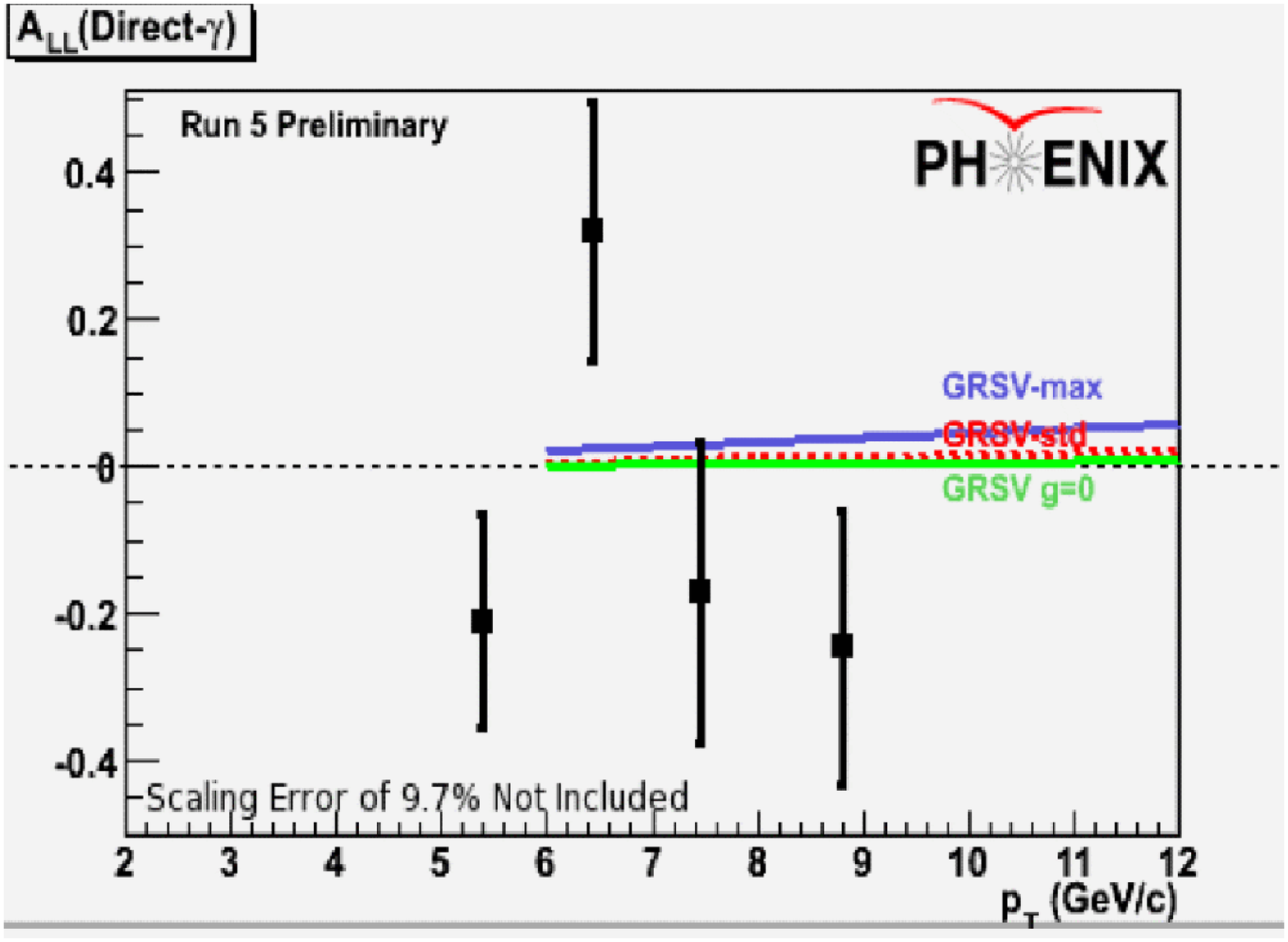}
\caption{(left)$e^{\pm} A_{LL}(p_{T})$. (right) Direct $\gamma$ $A_{LL}(p_{T})$ at $\sqrt{s}$=200GeV}
\label{fig6}
\end{figure} 

\section{Transverse Spin Measurements}
\begin{figure}[h]
\centering 
\includegraphics[scale=0.24]{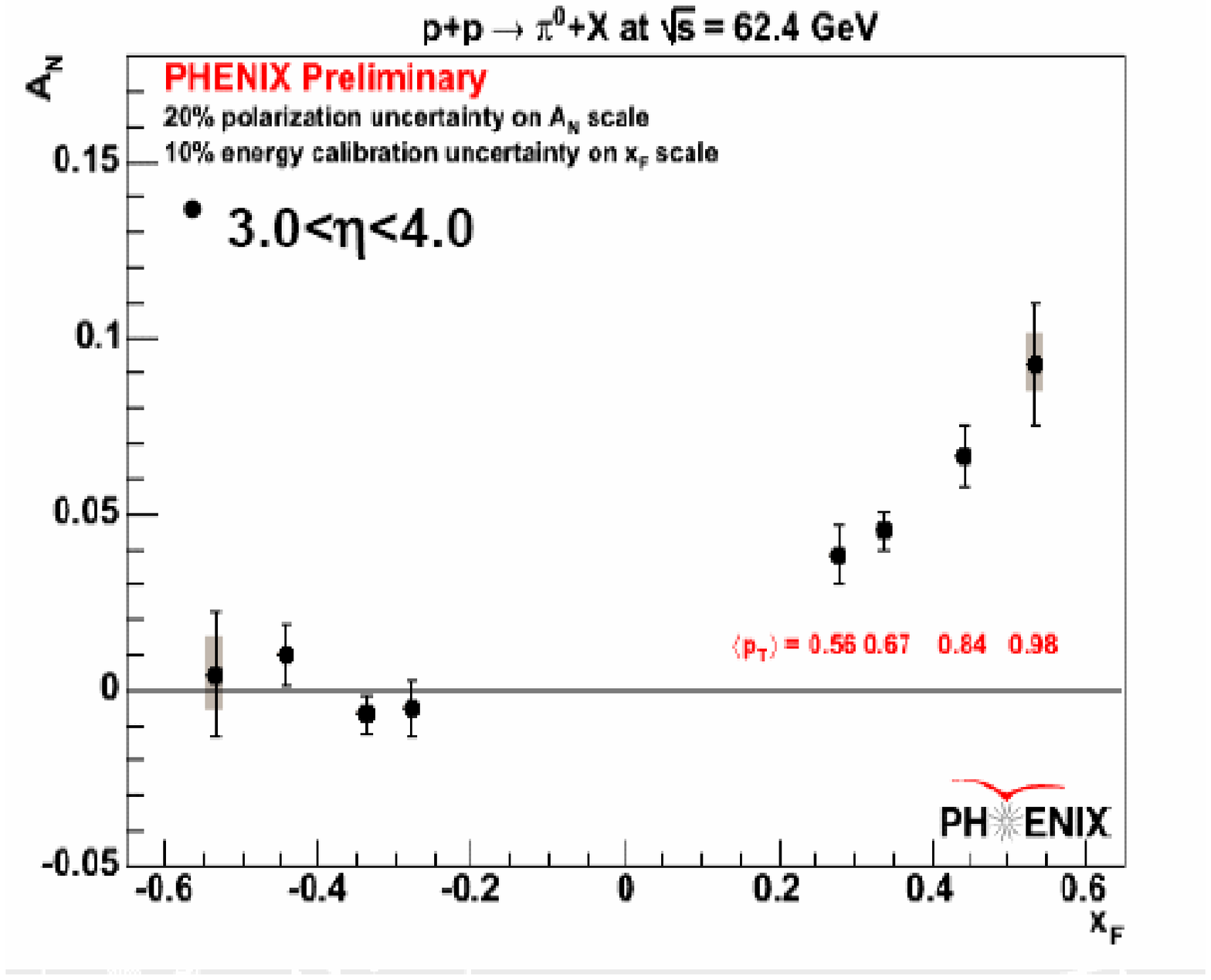}\includegraphics[scale=0.39]{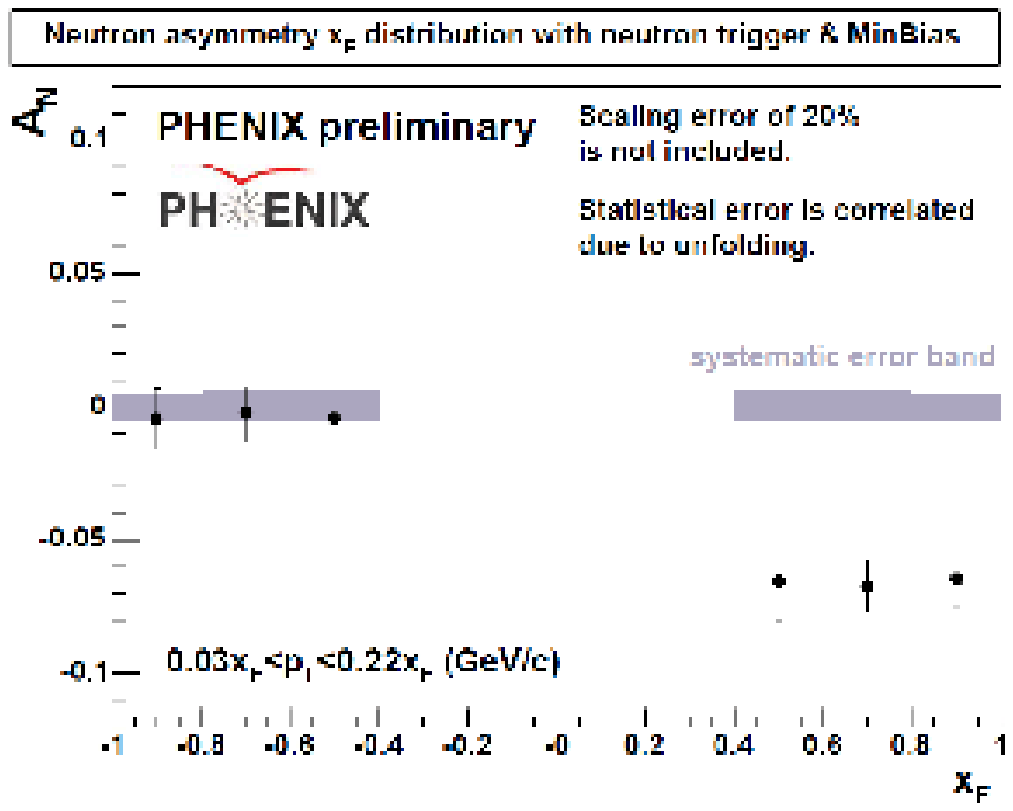}
\caption{$A_{N}$ of $\pi^{0}$ at $\sqrt{s}$=62.4 GeV and forward $X_{F}$. (right)Neutron $A_{N}$ vs $X_{F}$}
\label{fig7}
\end{figure}

PHENIX has measured small single spin asymmetries (SSA) $A_{N}$ of $\pi^{0}$ and $h^{\pm}$ at small $p_{T}$ for $|\eta| <0.35 $ and has helped constrain the magnitude of the gluon Sivers function[8]. In contrast, measured $A_{N}$ at forward $X_{F}$ show large asymmetries in the the positive but not the negative $X_{F}$ region. These interesting measurements may provide quantitative tests for theories involving valence quark effects. Other measurements include the $A_{N}$ of $J/\Psi$ which may also be sensitive to g-Sivers via D Meson production as its produced from g-g fusion[10], neutron's $A_{N}$(Fig.\ref{fig7}), and $k_{T}$ asymmetries which aim at probing orbital angular momentum[9]
\section{Conclusions}\label{concl}
PHENIX is well suited to the study of spin structure of the proton with a wide variety of probes. A variety of new results aiming to disentangle spin partonic contributions are emerging. In the next coming years, statistics needed to explore different channels for different gluon kinematics and different mixtures of subprocesses will become available and allow a more accurate picture of the spin of the proton.

\small
\vfill\eject

\begin{thebibliography}{11}  
\small{
\bibitem{bib1}Ashman J, et al {\it Phys. Lett.} {\bf B206}(1988) 364

\bibitem{bib2}G. Bunce et al., {\it Annu. Rev. Nucl. Part. Sci.} {\bf 50} (2000) 525.
  
\bibitem{bib3}Alekseev I, et al. {\it Design Manual Polarized RHIC} $http://www.agsrhichome.bnl.gov/RHIC/Spin/design/$ (1998)
  
\bibitem{bib4}K. Adcox et al., {\it Nucl. Inst. Meth. A} {\bf 499} (2003) 469.

\bibitem{bib5}D. de Florian et al, ArXiv:0804.0422 (2008)

\bibitem{bib6}M. Gl$\ddot{u}$ck et al, Phys. Rev. D 63, 094005 (2001).

\bibitem{bib7}T. Gehrmann et al,  Phys. Rev. D 53, 6100 (1996).

\bibitem{bib8}Anselmino et al.,  Phys. Rev. D 74, 094011 (2006).

\bibitem{bib9}Adler et al., Phys. Rev. D 74, 072002 (2006).

\bibitem{bib10}Anselmino et al., Phys Rev D 70, 074025 (2004).

\bibitem{bib11}Adare et al., Phys. Rev. D 76, 051106 (2007)

}
\end{thebibliography}
\end{document}